\documentclass[10pt]{article}
\usepackage{amsmath,amssymb}
\usepackage{graphics}
\begin{document}
\topmargin=6pt
\evensidemargin .2in
\oddsidemargin .2in
\baselineskip=6pt
\begin{center}
{\Large{\bf   Empirical Determination of Threshold Partial Wave Amplitudes in $p\,p \to p\,p\,\omega$
 }}\\
\vspace{0.5cm}
                                                                                
{\bf  G. Ramachandran$^1$, J. Balasubramanyam$^{2,3}$, M. S. Vidya$^4$, and  Venkataraya$^5$. }\\
{\it $^1$ Indian Institute of Astrophysics,Koramangala,Bangalore,560034}\\{\it $^2$K. S. Institute of Technology, %%@
Bangalore, 560062,India } \\
{\it $^3$ Department of Physics, Bangalore University, Bangalore, 560034,India. }\\
{\it $^4$ V/17, NCERT Campus, Sri Aurobindo Marg, New Delhi 110016, India. }\\
{\it $^5$ Vijaya College, Bangalore, 560011, India. }
\end{center}

\begin{abstract}
Using the model independent irreducible tensor approach to $\omega$ production
in $pp$ collisions, we show theoretically that, it is advantageous to measure 
experimentally the polarization of $\omega$, in addition to the proposed 
experimental study employing a polarized beam and a polarized target.
\end{abstract}

Threshold production of light as well as heavy mesons in $NN$ collisions has 
attracted considerable attention \cite{mac,naka} in recent years, as the reactions 
are sensitive to the short range $NN$ interaction and involve only a few partial 
waves. Experimental studies have reached a high degree of sophistication with 
measurements of spin observables in charged \cite{prez} as well as neutral 
\cite{mey} pion production in $\vec{p}\vec{p}$ collisions. Amongst the various 
proposed theoretical models including those which invoke subnucleonic degrees of 
freedom, the Julich meson exchange model \cite{han} may be said to have yielded 
theoretical predictions which are nearer to data. Although this  model 
was more successful in the case of charged pions \cite{prez}, it failed to provide an overall 
satisfactory reproduction of the data on neutral pions \cite{mey}. Both  Moskal 
et al\cite{mac}., and Hanhart \cite{mac},  have  remarked that 
`` apart from rare cases, it is difficult to extract a particular piece of information 
from the data". In this context, a model independent approach \cite{gr1} which was 
developed using irreducible tensor techniques \cite{gr2} has been employed 
\cite{pnd1} to analyze the data \cite{mey} on  $\vec{p}\vec{p} \to pp\pi^0$ and
Deepak, Haidenbauer and Hanhart \cite{pnd1} have recently found that the Julich model deviates  
very strongly from empirically extracted estimates  for the $^3P_1\to\, ^3P_0p$ 
and to a lesser extent for the $^3F_3\to\, ^3P_2p$. They also find that the 
$\Delta$ degree of freedom is important for the quantitative   understanding 
of the reaction. The analysis has reiterated once again the importance of $\Delta$
contribution which has been noted in several earlier studies \cite{bkj}. 
The rich spin structure \cite{sil} of $NN \to N\Delta$ and $ N \Delta \to N \Delta$ 
has also been analyzed. Of the sixteen amplitudes associated with $NN \to N\Delta$
as many as ten are second rank spin tensors and Ray \cite{ray} has drawn attention
to their importance based on a partial wave expansion model where he found that 
``the total and differential cross-section reduced by about one half, the structure 
in the analyzing powers increased dramatically, the predictions of $D_{NN}$ 
became much too negative, while that for $D_{LL}$ became much too positive and 
the spin correlation predictions were much too small when all ten of the rank 2
tensor amplitudes were set to zero, while the remaining six amplitudes were 
unchanged.'' The study of meson production has also focussed attention on the ``missing
resonance problem" \cite{bar} which refers to the predicted \cite{add} highly excited $N^*$ states which
have not been seen in $\pi N$ scattering. Moreover heavy meson production not only
probes distances \cite{naka} which are shorter than that in the case of pion production, but 
also the strange quark content of the nucleon. In particular  the cross-section
ratio for $NN \to NN \omega / \phi$ has been measured \cite{bal} in view of the 
dramatic violations \cite{ams} of the OZI \cite{oku} rule observed in $\bar{p}p$ 
collisions. Heavy meson production has also attracted attention in the context 
of dilepton spectra and medium modifications \cite{fes}. In particular  the total 
 cross-sections for $pp \to pp \omega$ have been measured \cite{heb}at 
five c.m. energies in the range 3.8 MeV to 30 MeV above threshold and the 
total and differential cross-sections  \cite{abd} at 92 and
173 MeV above threshold. There is also a proposal \cite{rat} to study experimentally 
the heavy meson production in $\vec{N}\vec{N}$ collisions.  A model independent 
irreducible tensor formalism \cite{gr3} has recently been developed to analyze 
such measurements where it was also pointed out that the polarization of $\omega$
can be studied by looking at the decay $\omega \to \pi^0 \gamma$. 

The purpose of the present paper is to point out that  measurements of the 
differential cross-section together with the spin polarization of $\omega$ and 
the analyzing powers are sufficient to determine empirically the leading partial 
wave amplitudes at threshold without any discrete ambiguities. The partial wave
amplitudes not only depend  on the c.m. energy $E$ at which the reaction 
takes place but also on the invariant mass $W$ give in natural units by
\begin{equation}
\label{invmass}
W = (E^2 + M_{\omega} - 2EE_{\omega})^{1/2},
\end{equation}
of the two protons in the final state, where $E_{\omega}$ denotes the energy of
the meson and $M_{\omega}$ its rest mass. If ${\boldsymbol p}_i$ and 
${\boldsymbol p}_f$ denote respectively the initial and final relative momenta between
the two protons in their respective c.m frames, we have
\begin{equation}
E^2 = 4 (p_i^2 + M^2),\, W^2 = 4(p_f^2 + M^2),
\end{equation}
where $M$ denotes the rest mass of the proton. Choosing $W$ (or equivalently 
$E_{\omega}$)and the polar angles $(\theta_f, \varphi_f)$ of ${\boldsymbol p}_f$ 
together with the polar angles $(\theta,\varphi)$ of the meson momentum ${\boldsymbol q}$ 
in the c.m frame as the five independent kinematical variables, we may write the 
unpolarized differential cross section, as 
\begin{eqnarray}
\label{updc}
\frac{d^5 \sigma_0}{dW d\Omega d\Omega}_f &=& 
(2 \pi)^{-5}\frac{W\,E_\omega\,(E-E_\omega)}{16\,p_i}\,q\,p_f {\tt{T}}r(TT^\dagger)
\nonumber \\
&=&  \frac{1}{4}{\tt T}r({\mathcal M}{\mathcal M}^\dagger),
\end{eqnarray} 
in a kinematically complete  experiment, where $T$  denotes the on-energy-shell
transition-matrix for the reaction, $T^{\dagger}$ its hermitian conjugate  and 
${\tt T}r$ denotes trace. Following \cite{gr2} we may express ${\mathcal M}$,
in a model independent way, as 
\begin{equation}
\label{tm}
\mathcal{M} = \sum_{s_f,s_i=0}^1 \sum_{\lambda=|s_f-s_i|}^{(s_f+s_i)}
\sum_{S=|1-s_f|}^{(1+s_f)}\sum_{\Lambda = |S-s_i|}^{(S+s_i)}
(( S^1(1,0) \otimes S^{\lambda}(s_f,s_i))^{\Lambda} \cdot 
\mathcal{M}^{\Lambda}(Ss_fs_i;\lambda)),
\end{equation}

where the irreducible tensor amplitudes $\mathcal{M}^{\Lambda}_{\nu}(Ss_fs_i;\lambda)$
of rank $\Lambda$ are given by
\begin{eqnarray}
\label{TLnu}
\mathcal{M}^{\Lambda}_\nu(Ss_fs_i;\lambda)&=&W(1 s_f \Lambda s_i ;S
\lambda)[\lambda]\sum_{j,ll_fLl_i}  f_{Ss_fs_i,ll_fLl_i}^{j}
  \,W(s_il_iSL;j \Lambda) \nonumber \\
&\times & (( Y_l(\boldsymbol{\hat{q}})\otimes
Y_{l_f}(\boldsymbol{\hat{p}}_f))^L   \otimes
Y_{l_i}(\boldsymbol{\hat{p}}_i))_{\nu}^{\Lambda},
\end{eqnarray}
where $(l_i,s_i)$ and $(l_f,s_f)$ characterize respectively initial and final states
of the $NN$ system in terms of their relative orbital angular momentum and total
spin quantum numbers, $j$ denotes the total angular momentum which is conserved
and $l$, the orbital angular momentum of the emitted spin 1 meson. 
The channel spin quantum number $S$ in the final state is the resultant of  
combining the  spin 1 of the $\omega$ with $s_f$ and likewise $l$ and $l_f$ combine 
to give $L$.The partial wave amplitudes
\begin{eqnarray}
\label{pwa}
f_{Ss_fs_i,ll_fLl_i}^{j} & = &(4 \pi)^{-2} (-1)^{L+l_i+s_i-j} 
[j]^2[S][1]^{-1}[s_f]^{-1} \nonumber \\  & & \times
\langle((ll_f)L(1s_f)S)j||T||(l_is_i)j \rangle,
\end{eqnarray}
depend only on the c.m. energy E, and invariant mass $W$ of the two nucleon system
in the final state. The above equations are  valid for  all c.m energies $E$.
In particular, at threshold, we may set $l_f=0$. In view of the observed \cite{abd} 
anisotropic angular distribution of $\omega$ at 173 MeV excess energy above 
threshold, we may take into consideration both $l=0$ and $1$. We then have to 
consider only two irreducible tensor amplitudes 
\begin{eqnarray}
\label{ita1}
{\mathcal M}^1_{\nu}(101;1) &=& (12\sqrt{3}\pi)^{-1}Y_{1\nu}( \hat{{\boldsymbol p}}_i)f_1 \\
\label{ita2}
{\mathcal M}^1_{\nu}(100;0) &=& (12\pi)^{-1}Y_{1\nu}( \hat{{\boldsymbol q}})f_2 
 + (6\sqrt{5\pi})^{-1}
(Y_1(\hat{{\boldsymbol q}})\otimes Y_2( \hat{{\boldsymbol p}}_i))^1_{\nu}f_3,
\end{eqnarray}
where the short hand notation $f_1,f_2,f_3$ is used for convenience to denote threshold
partial wave amplitudes shown in Table I. After integration with respect to 
$d\Omega_{p_f}$, the unpolarized differential cross-section is obtained as

\begin{table}[t]
\begin{center} 
\caption{\label{amplitudes} Threshold partial wave amplitudes for 
$pp \to pp \omega$.}
\begin{tabular}{lcr}\\ \hline \hline
Partial Wave & Initial State & Final State \\
Amplitudes &  &  \\ \hline
$f_1 = f_{101; 0001}^{1}$ \, & \, $^3P_1$ \, & \, $ (^1Ss)\,^3\mathcal{S}_1 $ \\
$f_2 = f_{100; 1010}^{0}$ & $^1S_0$ & $ (^1Sp)\,^3\mathcal{P}_0 $ \\
$f_3 = f_{100; 1012}^{2}$ & $^3D_2$ & $ (^1Sp)\,^3\mathcal{P}_2 $ \\ \hline \hline
\end{tabular}
\end{center}
\end{table}

\begin{equation}
\frac{d^3\sigma_0}{dW d\Omega}= \frac{1}{192\pi^2}
[a_0+ \frac{9}{10}\,a_2 \,cos^2 \theta],
\label{updcs}
\end{equation}
where an experimental measurement of (9) readily enable us to determine
the coefficients 

\begin{eqnarray}
\label{a}
a_0 & = & [|f_1|^2+3|f_2+\frac{1}{\sqrt{10}}f_3|^2] \\
a_2 & = & [|f_3|^2- 2\sqrt{10}\,\Re(f_2 f_{3}^{*})]
\end{eqnarray}

The state of polarization of $\omega$, with c.m energy $E_{\omega}$, may be defined 
in terms of its spin density matrix $\rho$ whose elements are given by
\begin{eqnarray}
\rho_{m m'} &=& \frac{1}{4}\sum_{s_f m_f} \int d\Omega_f 
\langle s_f m_f;1 m|\mathcal{M}\mathcal{M}^{\dagger}|1 m';s_f m_f \rangle\\
&=& \frac{{\tt T}r\rho}{3}\sum_{k=0}^{2} (-1)^q C(1k1;m'-qm)[k]t^k_q,
\end{eqnarray}
in terms of Fano statistical tensors $t^k_q$ of rank $k$ such that ${\tt T}r\rho$ is 
given by (9) and $t^0_0=1$. 

It is advantageous now to express the vector and tensor polarizations $t^1_q$ 
and $t^2_q$ in the transverse frame which is a  right handed frame whose Z-axis 
is chosen along ${\boldsymbol p}_i \times {\boldsymbol q}$ with ${\boldsymbol p}_i$ 
along X-axis. It may be noted that the polar angles of ${\boldsymbol q}$ in this 
frame are $(\frac{\pi}{2}, \theta)$ if we continue to use $\theta$ to denote the 
angle between ${\boldsymbol q}$ and ${\boldsymbol p}_i $, i.e.
${\boldsymbol q} \cdot {\boldsymbol p}_i = q\,p_i\,cos\theta$. We then have 
\begin{eqnarray}
\label{t11a}
{\tt T}r\rho \; t^1_0 &=&  \frac{3}{64 \pi^2}  \sqrt{\frac{3}{5}} \,b 
\,sin\theta \,cos\theta \\
\label{t20a}
{\tt T}r\rho \;t^2_0 &=& \frac{1}{384 \pi^2} \sqrt{\frac{1}{2}} \, [c_0 + 
\frac{18}{10} \,c_2 \,cos^2\theta]\\
\label{t22a}
{\tt T}r\rho \;t^2_{\pm 2} &=& \frac{1}{256 \pi^2}\sqrt{\frac{1}{3}}
\,[d_0 - 12 \,d_2\,cos^2\theta  \mp 6 \,i\, d_3\,sin2\theta]
\end{eqnarray}
in the transverse frame. The coefficients are given by

\begin{eqnarray}
\label{t11b}
b &=& -\Im (f_2f_3^*) \\
\label{t20bb}
c_0 &=& [-|f_1|^2+ 6|f_2+\frac{1}{\sqrt{10}}f_3|^2] \\ 
c_2 &=& a_2\\
d_0&=&  [|f_1|^2+ 6|f_2+\frac{1}{\sqrt{10}}f_3|^2] \\
\label{t21b}
d_2 &=& [|f_2|^2+ \frac{1}{4}|f_3|^2- \frac{1}{\sqrt{10}} \Re(f_2f_3^*)] \\ 
d_3 &=& [|f_2|^2-\frac{1}{5}|f_3|^2-\frac{1}{\sqrt{10}} \Re(f_2f_3^*)]
\end{eqnarray}

Thus the experimental measurement of differential cross-section (9) 
and $t^k_q$ enable us to determine empirically 

\begin{eqnarray}
\label{f1}
|f_1|^2 &=& \frac{1}{2}(d_0-c_0)\\
|f_2|^2 &=& \frac{1}{36}(d_0+c_0+8d_2+16d_3)\\
|f_3|^2 &=& \frac{20}{9}(d_2-d_3)\\
\Re (f_2f_3^*) &=& \frac{\sqrt{10}}{36}(d_0+c_0-8d_2-4d_3)\\
\Im (f_2f_3^*) &=& -b
\end{eqnarray}

Thus it is seen from our model independent theoretical analysis that priority 
should be given to measure the polarization of $\omega$ in $pp \to pp \vec{\omega} $. 
This enables us to determine empirically, not  only the strengths of the partial
wave amplitudes $f_1, f_2, f_3$ but also the relative phases between $f_2$ and 
$f_3$. Thus $|f_2+\frac{1}{\sqrt{10}}f_3|$ is known except for an overall phase.
The proposed study of $\vec{p}\vec{p} \to pp \omega$ can then be used to determine 
the relative phase between $(f_2+\frac{1}{\sqrt{10}}f_3)$ and  $f_1$ as follows.

If ${\boldsymbol P}$ and ${\boldsymbol Q}$ denote respectively the beam and target 
polarizations, the differential cross-section for $\vec{p}\vec{p} \to pp\omega$ 
is given by

\begin{equation}
\frac{d^3\sigma}{dW d\Omega} = \int d\Omega_{p_f}
{\tt T}r({\mathcal M}\rho^i{\mathcal M}^\dagger)
\end{equation}
where
\begin{equation}
\rho^i = {\textstyle{\frac{1}{4}}}(1+\boldsymbol{\sigma}_1
\cdot \boldsymbol{P})(1+\boldsymbol{\sigma}_2 \cdot \boldsymbol{Q}),
\end{equation}

This leads to \cite{gr3} 

\begin{equation}
\frac{d^3\sigma}{dW d\Omega} = 
\frac{d^3\sigma_0}{dW d\Omega}
[1 + {\boldsymbol P} \cdot {\boldsymbol A}^B + {\boldsymbol Q} \cdot {\boldsymbol A}^T
+ \sum_{k=0}^{2}((P^1 \otimes Q^1)^k \cdot A^k)]
\end{equation}

The vector analyzing powers ${\boldsymbol A}^B,{\boldsymbol A}^T$ and ${\boldsymbol A}$
are normal to the reaction plane containing ${\boldsymbol p}_i$ and ${\boldsymbol q}$.
Thus

\begin{eqnarray}
\label{ana}
A^B_Z - A^T_Z &=& \frac{1}{16 \pi^2} \sqrt{\frac{1}{6}}\,\Im [f_1(f_2+\frac{1}{\sqrt{10}}f_3)^*]
sin \theta, \\ 
\label{cor}
A_Z &=& \frac{i}{32 \pi^2} \sqrt{\frac{1}{3}} \,\Re [f_1(f_2+\frac{1}{\sqrt{10}}f_3)^*]sin \theta
\end{eqnarray}
in the transverse frame.

Since $|f_1|$ is known from \eqref{f1} the relative phase between $f_1$ and  $(f_2+\frac{1}{\sqrt{10}}f_3)$
is determinable without any trigonometric ambiguity from \eqref{ana} and \eqref{cor}.
Thus $f_1, f_2$ and $f_3$ are determinable empirically except for an overall phase.

In summary, therefore, we advocate measurement of polarization of $\omega$ in 
$pp \to pp \vec{\omega}$ in addition to the proposed experiments \cite{rat} on 
$\vec{p}\vec{p} \to pp \omega$,  as this will enable  the complete empirical 
determination of the leading threshold amplitudes $f_1, f_2$ and $f_3$ without 
any discrete ambiguities.

\section*{Acknowledgments}
One of us (G.R.) is grateful to Professors B.V. Sreekantan, R. Cowsik,
J.H. Sastry and R. Srinivasan for facilities provided for research at the 
Indian Institute of Astrophysics and another (J.B.) acknowledges much
encouragement for research given by the Principal Dr. T.G.S. Moorthy and the
Management of K.S. Institute of Technology.

\end{document}